\input epsf.sty
\documentstyle[12pt]{article}
\newcommand{\rf}{\par\noindent\hangindent 1cm {}}
\def\pks{PKS~2155--304\,}
\begin{document}
\def\cl{\centerline}
\def\ref{\par\noindent\hangindent 20pt}
\def\mincir{\raise -2.truept\hbox{\rlap{\hbox{$\sim$}}\raise5.truept
\hbox{$<$}\ }}
\def\magcir{\raise -2.truept\hbox{\rlap{\hbox{$\sim$}}\raise5.truept
\hbox{$>$}\ }}
\Large
\begin{center}
{\bf
Intensive optical monitoring of the BL Lac object PKS 2155-304:
the July 1998 campaign 
}
\end{center}
\bigskip
\normalsize
\begin{center}
{\bf
E. Poretti$^1$, 
E. Pian$^2$, 
A. Treves$^3$ 
}
\bigskip

{\small
{\sl

$^1$ Osservatorio di Brera, Via E. Bianchi 46, I-23807 Merate (LC), Italy
-- email: poretti@merate.mi.astro.it\\
$^2$ ITESRE, CNR, Via Gobetti 101, I-40129, Bologna, Italy -- email:
pian@tesre.bo.cnr.it\\
$^3$ University of Como, Via Lucini 3, I-22100 Como, Italy -- email:
treves@uni.mi.astro.it \\
}}

\end{center}

\bigskip\noindent
{\bf Abstract. }
\bigskip

Fast blazar variability on interday and intraday time scales is an
extremely powerful tool to investigate the structure of the AGN emitting
regions and the nature of the processes responsible for energy production.
Since PKS 2155--304 is the brightest BL Lac at the optical and UV
wavelengths, we have undertaken a program of systematic intensive optical
monitoring of this blazar in search of fast variability events.  We
present here the results of a CCD photometric campaign 
conducted at ESO (La Silla, Chile) on 6-14 July 1998, for a total time
on the source of $\sim$ 30 hours. 
\bigskip

{\bf 1. Introduction}

\bigskip

\noindent
Optical fast variations in blazars could be the signature of physical or
geometrical variations within the jet on small scales, and therefore can
trace the propagation of disturbances and help in mapping the emission
regions in detail.  PKS~2155--304 is the brightest and best monitored BL
Lac at the optical and UV wavelengths. Significant variability in this
spectral region on various time scales (from days to few hours) has been
commonly observed in this source (Smith \& Sitko 1991; Carini \& Miller
1992; Smith et al. 1992; Urry et al. 1993; Courvoisier et al. 1995; Heidt
et al. 1997; Bai et al. 1998).  

Some studies (Paltani et al. 1997, Pesce et al. 1997, Pian et al.
1997) have shown that rapid (time scales of $\sim$1 hour or less) optical
and UV flux variability events can occur. Spectral
variability is modest, and weakly or not clearly correlated with flux.  
When a correlation is found, this suggests flatter spectral shapes during
brighter states, in qualitative agreement with models based on radiative
cooling (Paltani et al. 1997).

Well and regularly sampled optical light curves on long time intervals are
necessary to apply the correlation methods in search of typical time
scales. Therefore, we have started a systematic program of fast
variability monitoring of PKS 2155--304 specifically aimed at sampling its
rapid optical variations.  Since the optical spectral continuum is likely
produced in the inner jet, variability in this spectral
region provides the clearest evidence for dynamic processes occurring near
the active nucleus.  Results from campaigns conducted as part of this
project at ESO and Las Campanas telescopes in 1996 and 1997 have been
presented by Mantegazza et al. (1999).

Here we report on the photometric monitoring of PKS~2155--304 in July 1998. 
\vfill
\eject
{\bf 
2. Observations and data reduction procedure}
\bigskip

\noindent
CCD photometry of \pks was obtained with the ESO Dutch 0.91--m telescope 
in 8 consecutive nights from July 6 to 14, 1998. 
The CCD detector was the ESO N~33, a 512 pixels~x~512 pixels TEK chip.
\pks and the two field stars measured by Smith et al. (1991, see their
Tab.3 where
they are quoted as Number 2 and 3) could be included in the 3.77x3.77 arcmin$^2$
field of view.  As in classical
differential photometry, we used the brighter star (i.e. the number 2) 
as the comparison one and the number 3 
as the check one.  We collected 351 images in $R$  and
331 in $V$  (see Tab. 1). 
Exposure times were set to have a high signal from \pks and comparison stars;
usually they are shorter than 60 sec.

{\small
\begin{table}[htbp]
\begin{center}
\begin{tabular}{|r|rr|c|rr|rr|rr|}
\hline
 & & &Net time on& \multicolumn{2}{c|}{Magnitudes of} & \multicolumn{4}{c|}{Standard deviations [mmag]}\\
Night & \multicolumn{2}{c|}{CCD Images} & \pks & \multicolumn
{2}{c|}{the check star} &\multicolumn{2}{c|}{Check star}&\multicolumn{2}{c|}{\pks}\\
\multicolumn{1}{|c|}{[JD]} & $V$ & $R$ & [hours] & \multicolumn{1}{c}{$V$} &
\multicolumn{1}{c|}{$R$}& $V$ & $R$&$V$&$R$\\
\hline
2451001 & 22 & 22 &  2.4 & 13.027 & 12.512 & 2.6 & 2.9 & 4.6 & 4.1 \\
2451002 & 57 & 74 &  5.3 & 13.023 & 12.503 & 5.4 & 3.4 & 6.2 &10.0 \\
2451003 & 8 & 11 &   0.2 & 13.022 & 12.508 & 2.4 & 1.5 & 5.3 & 5.1 \\
2451004 & 33 & 38 &  3.4 & 13.024 & 12.510 & 5.8 & 2.8 & 8.4 & 8.1 \\
2451005 & 21 & 20 &  7.2 & 13.023 & 12.505 & 5.5 & 3.5 &11.7 &13.1 \\
2451006 & 106 & 90 & 7.2 & 13.023 & 12.505 & 5.3 & 3.4 & 6.5 & 7.4 \\
2451007 & 44 & 44 &  1.9 & 13.025 & 12.510 & 3.9 & 2.7 & 3.5 & 4.3 \\
2451008 & 38 & 50 &  2.0 & 13.021 & 12.506 & 3.9 & 3.4 & 5.7 & 3.8 \\
\hline
\end{tabular}
\caption{\small Log of the optical 
observations from July 6, 1998 to July 14, 1998. $V$ and $R$ magnitudes 
refer to the measurements of the check star; standard deviations (in thousandths
of mag) both to the check star and to \pks.}
\end{center}
\end{table}
}
The reduction of the images was performed by using MIDAS package.
Since targets stars are bright, the aperture photometry technique could
be successfully applied. The flux was
calculated in a 12 arcsec circle centered on each object. Sky background was
subtracted by measuring local values; this strongly reduced the
effects of the moonlight.  Moreover, flat
fields on the sky were taken  at the beginning and at the
end of each night: as a result, the images were satisfactorily corrected and a very
stable photometric system could be maintained, as can be noted looking at
the magnitudes of the check star listed in Tab.~1. Magnitude differences were
transformed into $V$ and $R$ magnitudes by using $V$=12.07 and
$R$=11.69 for the comparison star, as determined on the basis of our previous
CCD campaigns. These values are in good agreement with the photoelectric 
magnitudes  reported by Smith et al. (1991), i.e. $V$=12.04 and $R$=11.64.

\begin{figure}[htbp]
\vspace{0cm}
\epsfxsize=8cm
\centerline{{\epsfbox{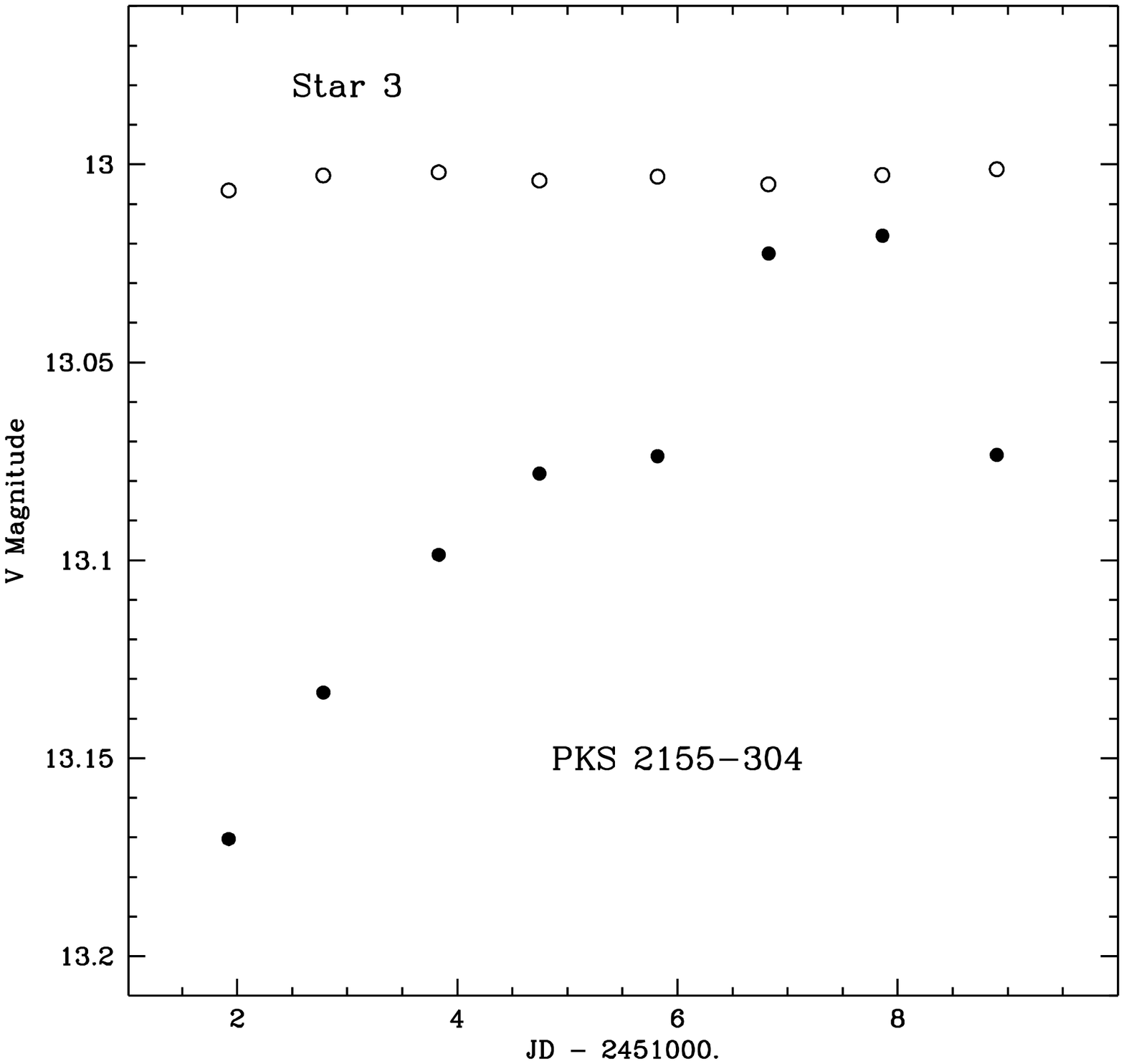}}{\epsfbox{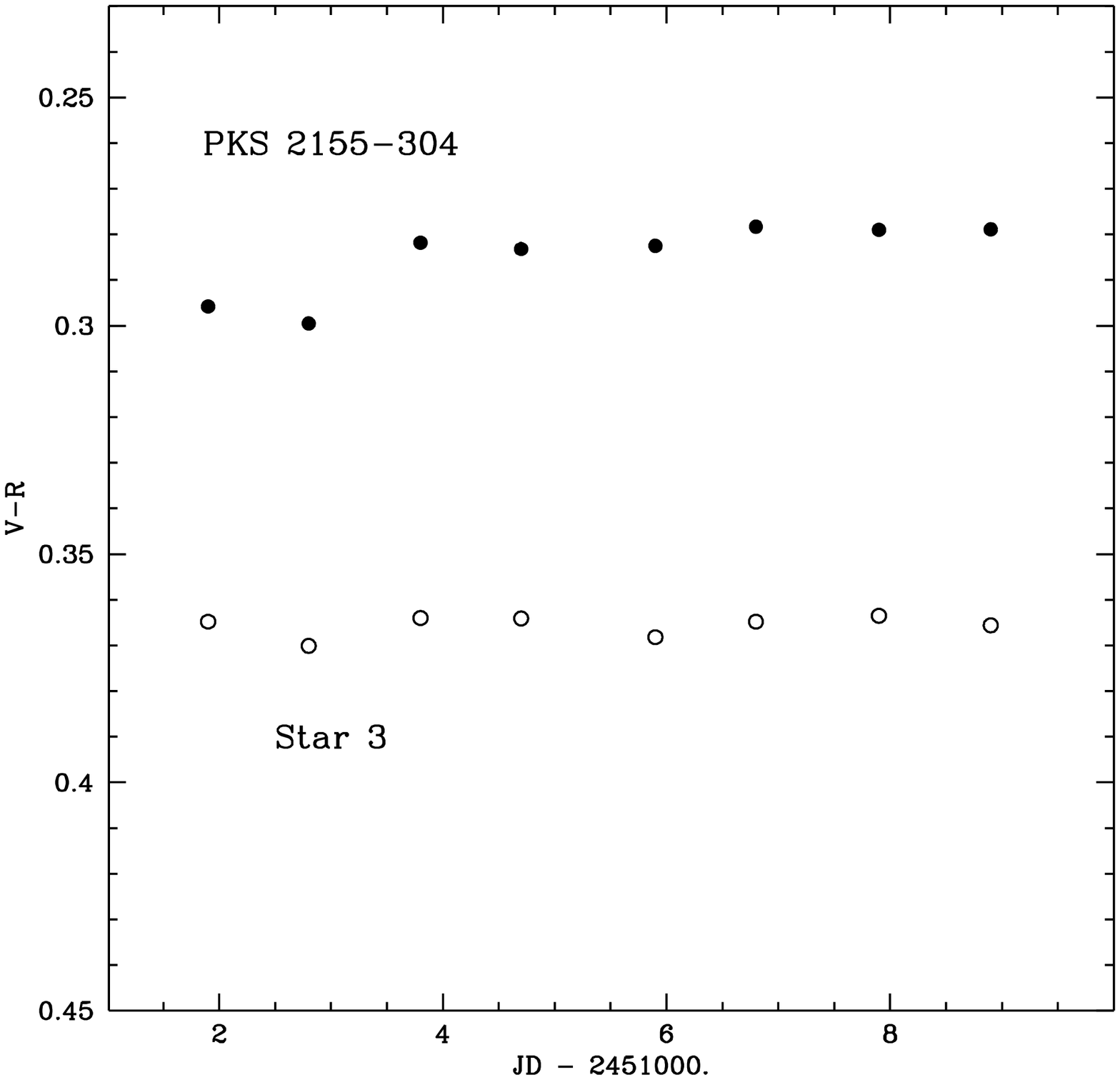}}}
\vspace{0cm}
\caption{\small 
The $V$ (left panel) and $V-R$ (right panel) curves of \pks (filled dots) as observed in
the ESO run in July 1998. $V$ magnitudes and $V-R$ colours of the check star (open
circles) are reported for comparison purposes.}
\end{figure}

\begin{figure}[htbp]
\vspace{0cm}
\centerline{\epsfxsize=10cm\epsfbox{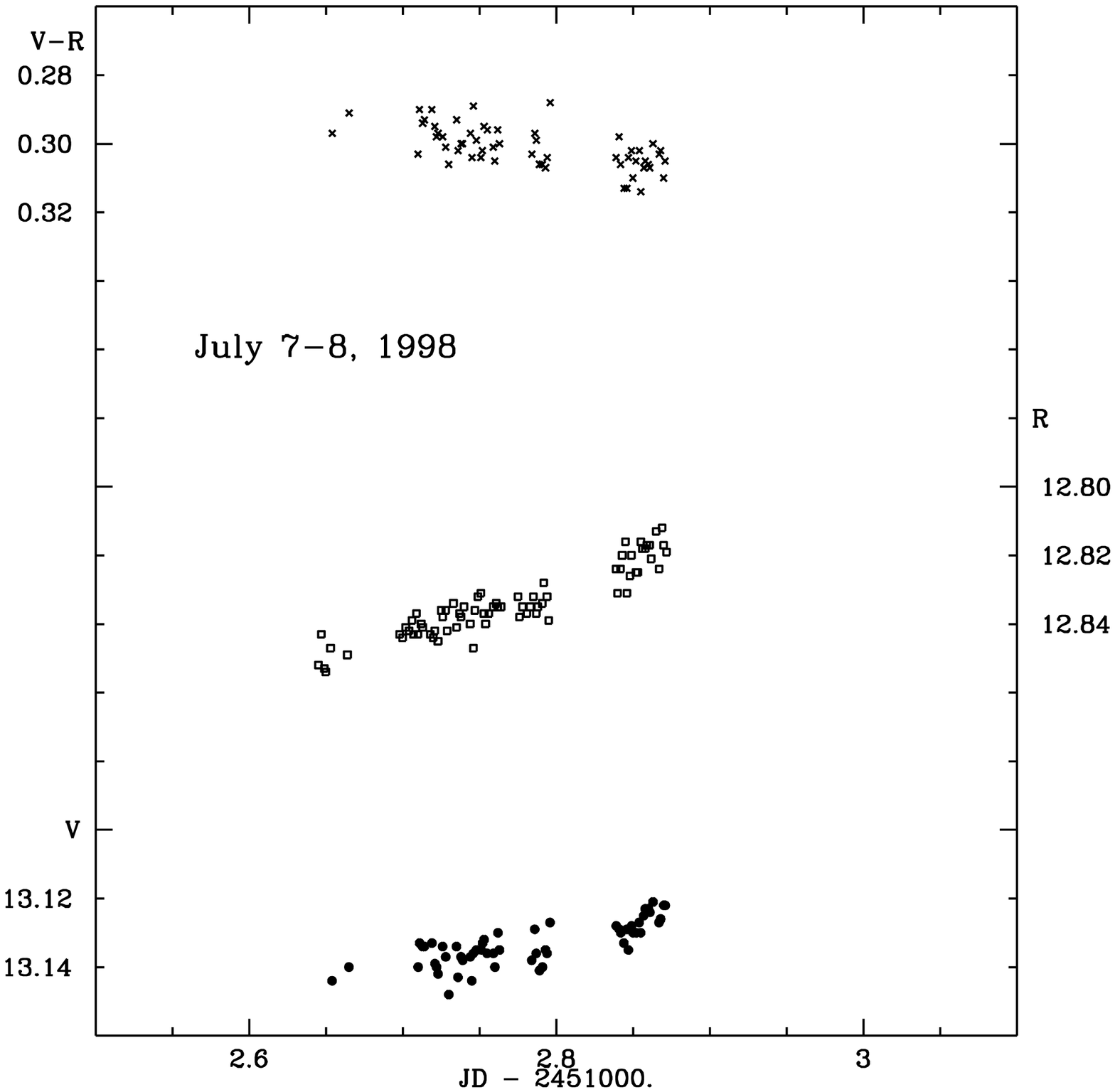}}
\vspace{0cm}
\caption{\small 
The $V$, $R$ and $V-R$ curves of \pks as observed in the second night
of our 1998 campaign.}
\end{figure}

\bigskip

{\bf 3. Photometric results}
\bigskip

\noindent
Figure 1 (left panel) shows the light curve in $V$ as obtained by averaging the magnitudes in
each night: a rather regular increase (0.15 mag) was observed during the
first seven nights, followed by a decline during the last night. The
simultaneous $V-R$ colour curve is shown in the right panel of Fig.~1: it is
 practically constant for most of time
(mean value around 0.28 mag), with an indication that it was somewhat 
redder ($\sim$0.30 mag) in the first two nights.

Intranight variability can be clearly noted by comparing the standard deviations of the
\pks measurements and those of the check star 
(see Tab.~1): they are greater by a factor $\sim$2 in the first five nights. 
Figure~2 shows the light curve observed during the night between July 7 and 8: a
continuous increase is observed. It should be noted 
that the brightening is larger in $R$--light than
in $V$--light, as appears in the $V-R$ colour curve. 
Such a reddening is quite unusual in the brightening
of a BL Lac object.

\bigskip

{\bf 4. Conclusions}

\bigskip
\noindent

The 1998 observations described here correspond to a state of the source slightly weaker
than in the 1997 Las Campanas campaign (Mantegazza et al. 1999): in $V$--light the range
 was 12.90--12.95
in 1997, while it was 13.02--13.17 in 1998. Moreover, the colour index $V-R$ was
slightly redder in 1998 (0.28--0.30 mag) than in 1997 (0.26--0.28 mag).

However, in correspondence of the highest state detected by us ($V$=12.83 in the ESO
observations performed in July 1997 just after the Las Campanas run), the $V-R$ colour
 curve was unexpectedly high
($V-R$=0.30). This fact, joined with the behaviour observed in the night of
7--8 July, 1998,  indicates that the magnitude--colour dependance is rather
complex,
being related to the intensity state of the source and to the temporal sampling.
\bigskip

{\bf References}

\bigskip
{\small
\rf
Bai J.M., Xie G.Z., Li K.H., Zhang X., Liu W.W., 1998, A\&AS 132, 83
\rf
Carini M. T., Miller H.R., 1992, ApJ 385, 146
\rf
Courvoisier T.J.-L., Blecha A., Bouchet P., et al., 1995, ApJ 438, 108
\rf
Heidt J., Wagner S.J., Wilhelm-Erkers U., 1997, A\&A 325, 27
\rf
Mantegazza L., Pian E., Poretti E., et al. 1999, in 
Takalo L.O., Sillanp\"{a}\"{a} A. (eds.)
ASP Conf. Ser. Vol. 159, {\it BL Lac Phenomenon}, p. 137
\rf
Paltani S., Courvoisier T.J.-L, Blecha A., Bratschi P., 1997, A\&A 327, 539
\rf
Pesce J.E., Urry C.M., Maraschi L., et al., 1997, ApJ 486, 770
\rf
Pian E., Urry C.M., Treves A., et al. 1997, ApJ 486, 784
\rf
Smith P.S., Sitko M.L. 1991, ApJ 383, 580 
\rf
Smith P.S., Jannuzi B.T., Elston R., 1991, ApJS 77, 67
\rf
Smith P.S., Hall P.B., Allen R.G., et al., 1992, ApJ 400, 115 
\rf
Urry C.M., Maraschi L., Edelson R., et al., 1993, ApJ 411, 614
\vfill
\eject

\Large
\begin{center}
{\bf
 Optical  Polarization of PKS 2155-304
}
\end{center}

\bigskip
\normalsize
\begin{center}
{\bf
A.\ Treves$^1$, E.\ Pian$^2$, F.\ Scaltriti$^3$, E.\ Palazzi$^2$, L.\
Tommasi$^4$
and E.\ Poretti$^5$
}

\bigskip
{\small
{\sl
$^1$
Universit\`a dell'Insubria, Dipartimento di Scienze, Via Lucini 3, I-22100
Como, Italy
-- email:
treves@uni.mi.astro.it\\

\smallskip
$^2$
Consiglio Nazionale delle Ricerche, Istituto di Tecnologie e Studio delle
Radiazioni
Extraterrestri, Via Gobetti 101, I-40129 Bologna, Italy
-- email:
pian@tesre.bo.cnr.it\\

\smallskip
$^3$
Osservatorio Astronomico di Torino, Strada Osservatorio 20, I-10025 Pino
Torinese (TO), Italy
-- email:
scaltriti@to.astro.it\\

\smallskip
$^4$
Universit\`a di Milano, Dipartimento di Fisica, Via Celoria 16, I-20133
Milano, Italy
-- email:
tommasi@ifctr.mi.cnr.it\\

\smallskip
$^5$
Osservatorio Astronomico di Brera, Via Bianchi 46, I-23807 Merate (LC),
Italy
-- email:
poretti@merate.mi.astro.it
}
}
\end{center}

\bigskip\noindent
{\bf Abstract. }
Linear polarization of the bright BL Lac object PKS 2155-304 was studied
in the
UBVRI bands with the Torino photopolarimeter at the
2.15 m CASLEO telescope. Observations were performed in June and August
1998
for a total of $\approx$45 hours. Between the two
epochs, the linear polarization in the V band increased from
$\approx$3.5\% to
$\approx$5.5\%,
and the position angle decreased from 100$^\circ$ to 70$^\circ$.
Significant day-to-day variations of both quantities are also observed.
An isolated "dip" of the polarization percentage
occurred between 17 and 19 June in all bands: a decrease and subsequent
rise of a
factor of
2 is seen in the linear polarization, accompanied
by a variation of the position angle by 90$^\circ$. Occasionally,
significant
intraday variations are seen, not exceeding 20\%. Further
monitoring in November, 1998 indicates that the blazar does not show
significant
circular polarization.

\bigskip
{\bf
1. Introduction
}

\bigskip\noindent
PKS 2155-304 is a bright closeby BL Lac object (V=12-13, z=0.116), which
has been studied
in the entire electromagnetic spectrum (e.g. Chiappetti et al. 1999).
Variability with time scales as short as 1000~s has been detected in the
UV
and X-ray bands (e.g. Urry et al. 1997).

Many polarimetric observations are available in literature,
in optical and UV bands (Smith \& Sitko 1991; Smith et al. 1992;
Allen et al. 1993;
Courvoisier et al. 1995; Pesce et al. 1997, Visvanathan \& Wills 1998).
All of them report variability both in
polarized flux (P) and position angle (PA), with approximate ranges $2\% <
\rm P < 13\%$ and
$85^\circ < \theta < 155^\circ$. Different values of PA are reported
in Visvanathan \& Wills (1998), who found 31$^\circ$ and 32$^\circ$,
in May and June 1979, respectively.
The maximum rates of variations
can be
roughly estimated in 1.5\%/day and 30$^\circ$/day. A small wavelength
dependence
is usually present in data, showing polarization increasing with
frequency,
while such a
clear dependence cannot be established for the position angle.

We present simultaneous UBVRI linear polarimetry
of this source, obtained in June 17-22 and August 26-28, 1998 from
Complejo
Astronomico el Leoncito (CASLEO, Argentina), for a total amount of 45
hours,
and circular
polarimetry, obtained in 1998 November 15,16, 18 and 19 for 1.1 hours in
total.

\begin{figure}[h]
\centerline{\epsfysize=13cm\epsfbox{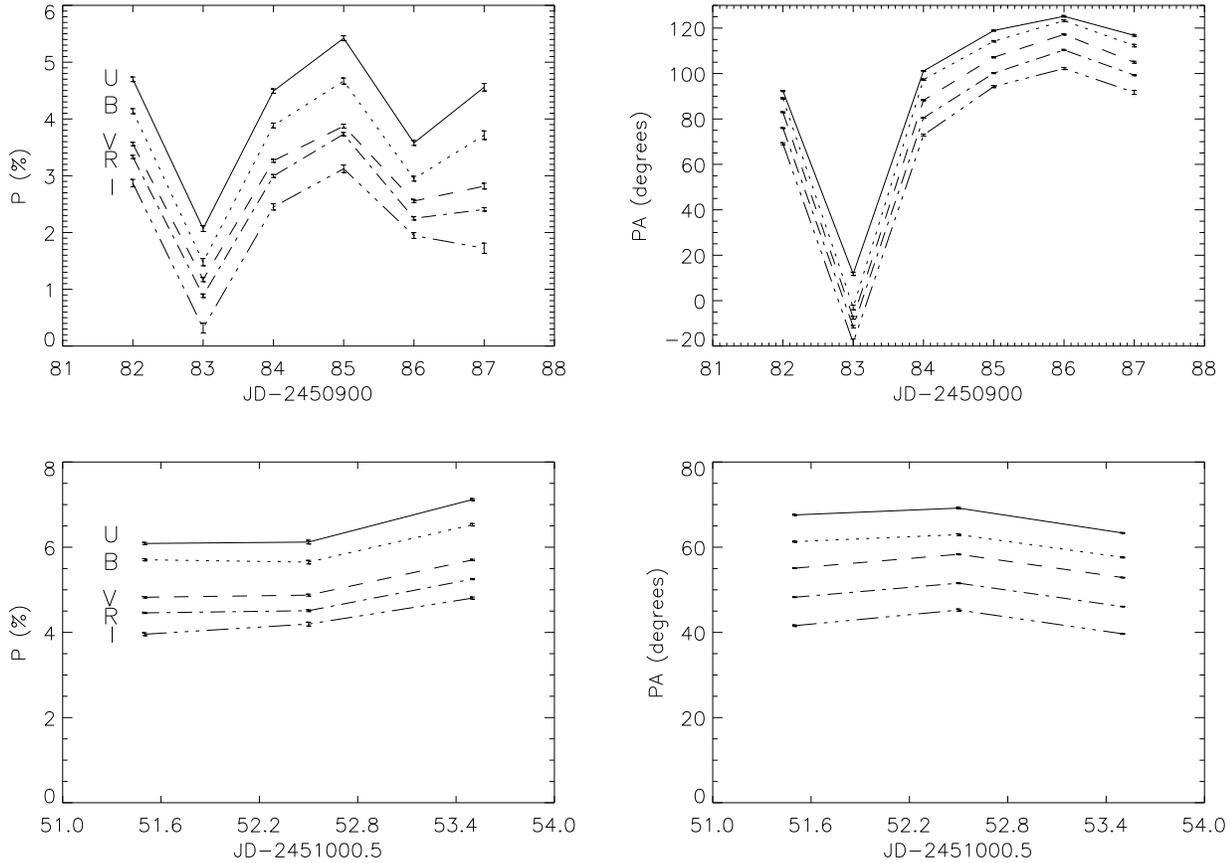}}
\caption{\small
Night weighted averages of linear polarization P and position angle PA in
the U, B, V, R, I bands
(1$\sigma$ error bars) for June 17-22 (upper two panels) and August 26-28,
1998. Y-axis
scales correspond to the U data. B, V, R, I are arbitrarily shifted down
with respect to U and to
each other by 0.3\% for P and 7$^\circ$ for PA. Lines have been added to
guide the eye only.
}
\label{plot21}
\end{figure}

\bigskip
{\bf
2. The observations
}

\bigskip\noindent
Observations have been performed with the 2.15~m telescope of CASLEO
Observatory using the Photopolarimetric System of Torino Observatory.
It has the capability to perform simultaneous observations
in UBVRI bands, thanks to four dichroic plus bandpass filters combinations
that
select
suitable bandpasses (equivalent to that of the standard UBVRI system) to
be analyzed by 5 different dedicated photomultipliers. A double
diaphragm rotating
wheel switches continuously between
the two images of the star (ordinary and extraordinary rays) produced by a
calcite slab
of suitable thickness. Both components of the sky background pass both
diaphragms and
polarization of the sky is directly eliminated.
The original design of the instrument can be found in Piirola (1973).
The integration time for each measurement, consisting of 8 positions of
the retarding plate, is 3.5 minutes. Observations were then binned
together in groups of four to improve statistics. The final time
resolution is about
15 minutes. Standard stars for PA calibration and instrumental
polarization subtraction
have been observed.

\bigskip
{\bf
3. Results
}

Fig.~3 reports the night weighted averages of linear
polarization together with
the position angle for June and August, 1998. The most noticeable feature
is the sharp decrease of P on the second night of June, corresponding
to a drastic change of PA. Fig.~4 shows a rather regular
variation of
P within one night in the U and B bands. Similar trends occur also in V,
R, I bands.
The dependence of P on the wavelength is illustrated in
Fig.~5.
Polarization decreases with wavelength with a flattening above 5000~\AA.

\begin{figure}
\centerline{\epsfysize=8.5cm\epsfbox{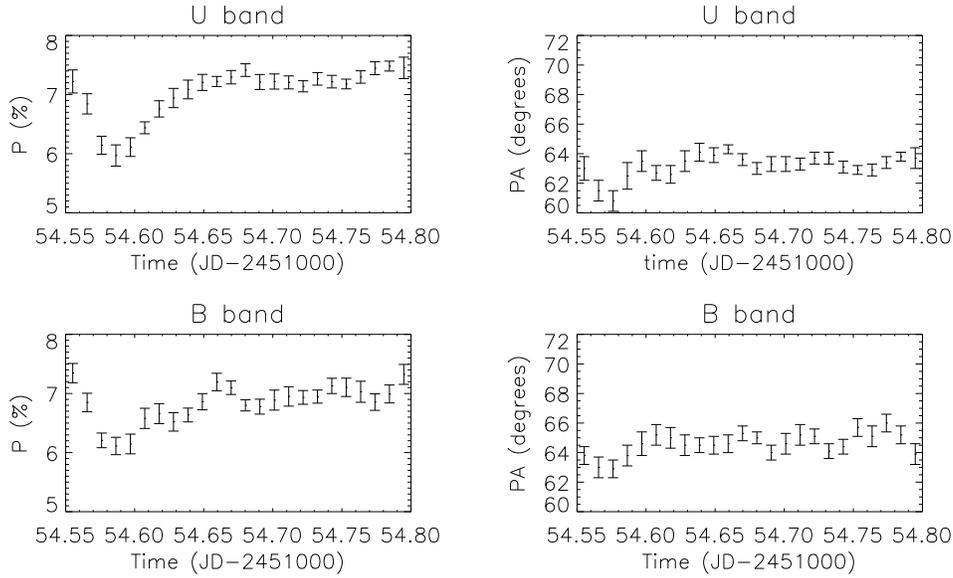}}
\caption{\small
A regular variation of P and PA recorded within the night of Aug. 28 in
the U and B band
(1$\sigma$ error bars).
}
\label{plot22}
\end{figure}

\begin{figure}[h]
\centerline{\epsfysize=6cm\epsfbox{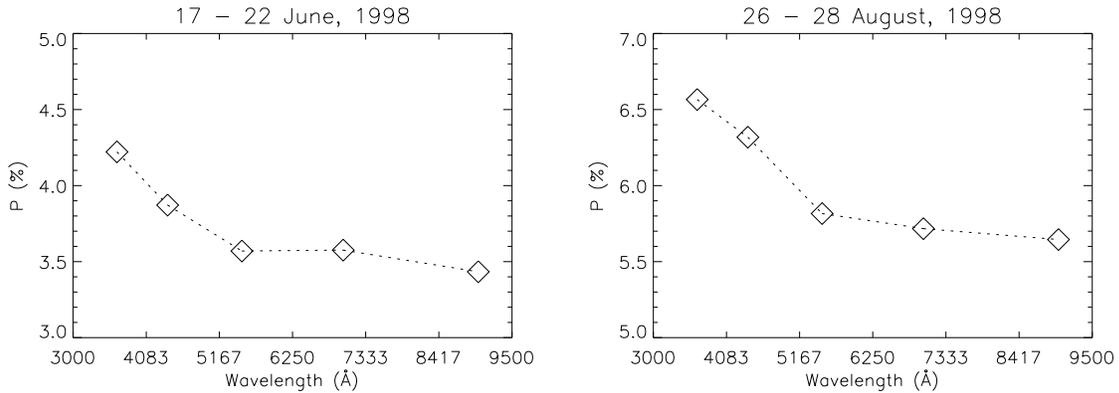}}
\caption{\small
Wavelength dependence of polarization.
Error bars are smaller than symbols used. Lines have been added to guide
the eye only.
}
\label{plot23}
\end{figure}

Circular polarization was searched for in November 15,16,18 and 19, 1998
with a total integration
time of 1.1 hours. Results are plotted in Fig.~6.
Upper limits in the five bands are $\approx$0.3 percent,
consistent with the results of Allen et al. (1993).

\begin{figure}[t]
\centerline{\epsfysize=11cm\epsfbox{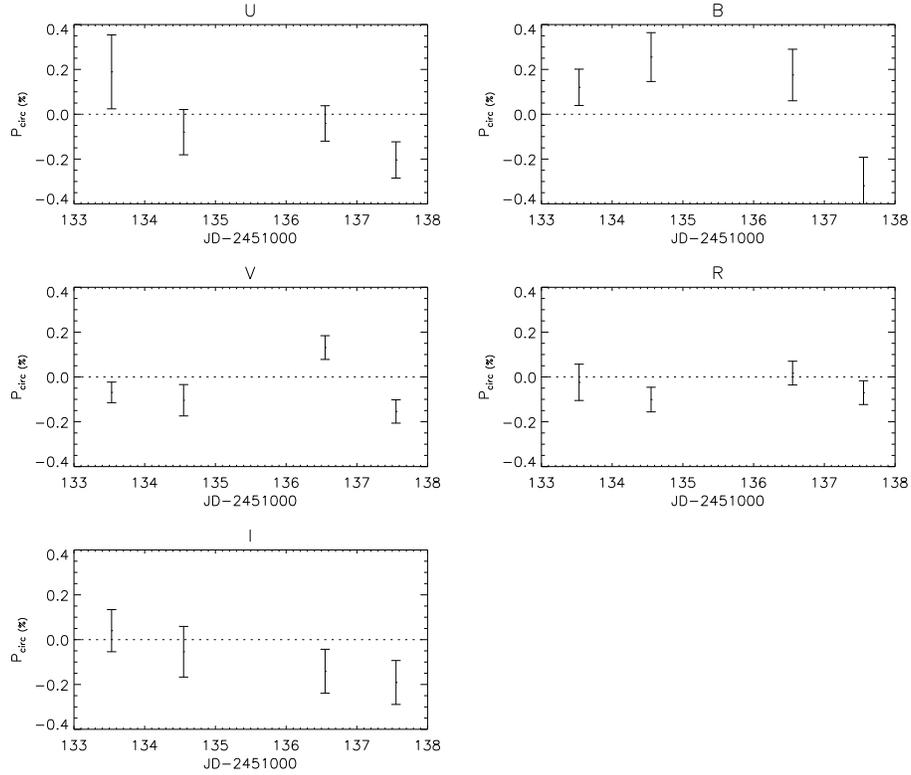}}
\caption{\small
Circular polarization vs. Julian Days in U, B, V, R, I bands in November
15-19, 1998.
}
\label{plot24}
\end{figure}


\bigskip
{\bf
4. Discussion
}

The variability and wavelength dependence indicate that the polarization
is mainly due to the BL Lac, rather than to the host galaxy, whose
magnitude, according to Falomo et al. (1991), is $m_I$=14.8.
The abrupt decrease of P in June 18, accompanied by a drastic variation of
PA,
suggests to us that the jet contains several components of different
polarization. The fading of one exalts the contribution of another.
This picture seems more natural than considering a modification of the
magnetic structure of the jet, i.e. an actual rotation of the field
orientation in
1-day time scale.

\bigskip
{\bf References}

\bigskip

{\small
\rf
Allen R.G., Smith P.S., Angel J.R.P. et al., 1993, ApJ 403, 610
\rf
Chiappetti L., Maraschi L., Tavecchio F. et al. 1999, ApJ 521, in press
\rf
Courvoisier T.G.L., Blecha A., Bouchet P. et al., 1995, ApJ 438, 108
\rf
Falomo R., Giraud E., Maraschi L. et al., 1991, ApJ 380, L67
\rf
Pesce J.E., Urry C.M., Maraschi L. et al., 1997, ApJ 486, 770
\rf
Piirola V., 1973, A\&A 27, 383
\rf
Smith P.S., Hall P.B., Allen R.G. et al., 1992, ApJ 400, 115
\rf
Smith P.S., Sitko M.L., 1991, ApJ 383, 580
\rf
Urry C.M., Treves A., Maraschi M. et al. 1997, ApJ 486, 799
\rf
Visvanathan N., Wills B.J., 1998, AJ 116, 2119
}

\end{document}